\documentclass{pasj00}

\begin{document}
\SetRunningHead{Doi et al.}{Bigradient phase referencing}
\Received{2006/1/2}
\Accepted{2006/4/26}

\title{Bigradient Phase Referencing}



%
 \author{
   Akihiro \textsc{Doi},\altaffilmark{1}
   Kenta \textsc{Fujisawa},\altaffilmark{1}
   Asao \textsc{Habe},\altaffilmark{2}
   Mareki \textsc{Honma},\altaffilmark{3,4}
   Noriyuki \textsc{Kawaguchi},\altaffilmark{3,4}\\ 
   Hideyuki \textsc{Kobayashi},\altaffilmark{5,3}
   Yasuhiro \textsc{Murata},\altaffilmark{6,7}
   Toshihiro \textsc{Omodaka},\altaffilmark{8}
   Hiroshi \textsc{Sudou},\altaffilmark{9}\\ 
   and
   Hiroshi \textsc{Takaba}\altaffilmark{9}}
 \altaffiltext{1}{Faculty of Science, Yamaguchi University, 1667-1 Yoshida, Yamaguchi, Yamaguchi 753-8512}
 \altaffiltext{2}{Division of Physics, Graduate School of Science, Hokkaido University, N10W8, Sapporo, Hokkaido 060-0810}
 \altaffiltext{3}{National Astronomical Observatory of Japan, 2-21-1 Osawa, Mitaka, Tokyo 181-8588}
 \altaffiltext{4}{Department of Astronomical Science, Graduate University for Advanced Studies,\\ 2-21-1 Osawa, Mitaka, Tokyo 181-8588}
 \altaffiltext{5}{Mizusawa VERA Observatory, 2-12 Hoshigaoka, Mizusawa, Oshu, Iwate 023-0861}
 \altaffiltext{6}{The Institute of Space and Astronautical Science, Japan Aerospace Exploration Agency,\\ 3-1-1 Yoshinodai, Sagamihara, Kanagawa 229-8510}
 \altaffiltext{7}{Department of Space and Astronautical Science, The Graduate University for Advanced Studies,\\ 3-1-1 Yoshinodai, Sagamihara, Kanagawa 229-8510}
 \altaffiltext{8}{Faculty of Science, Kagoshima University, 1-21-30 Korimoto, Kagoshima, Kagoshima 890-0065}
 \altaffiltext{9}{Faculty of Engineering, Gifu University, 1-1 Yanagido, Gifu 501-1193}

\KeyWords{astrometry --- atmospheric effects --- phase-referencing --- techniques: interferometric --- very-long-baseline interferometry} 

\maketitle

\begin{abstract}
We propose bigradient phase referencing (BPR), a new radio-observation technique, and report its performance using the Japanese very-long-baseline-interferometry network~(JVN).  In this method, a weak source is detected by phase-referencing using a primary calibrator, in order to play a role as a secondary calibrator for phase-referencing to a weak target.  We will be given the opportunity to select a calibrator from lots of milli-Jansky sources, one of which may be located at the position closer to the target.  With such a smaller separation, high-quality phase-referencing can be achieved.  Furthermore, a subsequent more-sophisticated calibration can relocate array's focus to a hypothetical point much closer to the target; a higher quality of phase referencing is available.  Our demonstrative observations with strong radio sources have proved the capabilities of BPR in terms of image dynamic ranges and astrometric reproducibility.  The image dynamic range on a target has been improved with a factor of about six compared to that of normal phase-referencing; the resultant position difference of target's emission between two epochs was only 62$\pm$50~micro-arcsecond, even with less than 2300-km baselines at 8.4~GHz and fast-switching of a target--calibrator pair of a 2.1-degree separation.
\end{abstract}

\section{Introduction}
A phase-referencing technique allows very-long-baseline interferometry~(VLBI) to make relative astrometry with an accuracy of less than 1~milli-arcsecond~(mas) and to detect very weak sources in mJy level.  The quality of phase referencing is limited by residual errors in differential phases between a target and a calibrator (e.g., \cite{Beasley&Conway1995}).  The most important error component is the uncertainty of atmospheric models in correlators.  The differential excess-path length between the two sources at different elevations is significantly harmful, even with a zenith phase-delay error of only a few centimeters and a separation of sources only one degree \citep{Beasley&Conway1995,Reid_etal.1999}.  Possible solutions are (1)~estimating the unknown phase-delay at zenith, (2)~to determine the residual phase gradient in the sky, and (3)~using a calibrator very close to a target.  It is advisable to apply all these solutions simultaneously.  The first solution can be achieved by geodetic-like observations \citep{Brunthaler_etal.2003,Mioduszewski&Kogan2004,Reid&Brunthaler2004} or by parallel plate air modeling of the long-term phase drifts \citep{Reid_etal.1999,Brunthaler_etal.2003}.  The data processing of the geodetic-like observations is supported by the task DELZN of the Astronomical Image Processing System~(AIPS; \cite{Greisen2003}), developed at the National Radio Astronomy Observatory.  The second solution can be achieved by using several strong calibrators over several degrees; this calibration process is also supported by the task ATMCA in the AIPS \citep{Fomalont&Kogan2005}.

The third solution, using a calibrator very close to a target, depends on a matter of blind chance.  However, it is promising if array's sensitivity is significantly improved, since the surface number density of radio sources dramatically increases in the fainter sky~(e.g., \cite{Fomalont_etal.1991}).  The present paper proposes a bigradient phase-referencing (BPR) method, which gives us a chance to utilize such weak radio sources as calibrators.  Once a weak source very close to a target is detected by phase-referencing using a strong calibrator, and then plays a role as a calibrator for phase-referencing to detect a very weak target.  Such a two-step approach is responsible for its name.  Furthermore, assuming that the residual phase components are linear around the target~\citep{Fomalont&Kopeikin2003}, array's focus on the weak calibrator can relocate to a point much closer to the target.  Hence, the BPR leads to be nearly free from the long-term phase drifts.

In the present paper, we present the BPR method and its observational tests.  In Section~\ref{section:principle} we describe the principle of the method and predict final phase errors.  In Section~\ref{section:testobs} we report demonstrative VLBI observations.  In Section~\ref{section:discussion} we discuss the capability of this method.  Finally, we summarize the method and the test observations in Section~\ref{section:summary}.

\section{Method}\label{section:principle}

\subsection{A calibrator arrangement and scheduling}\label{section:arrange&schedule}

\begin{figure}[hbt]
  \begin{center}
    \FigureFile(80mm,160mm){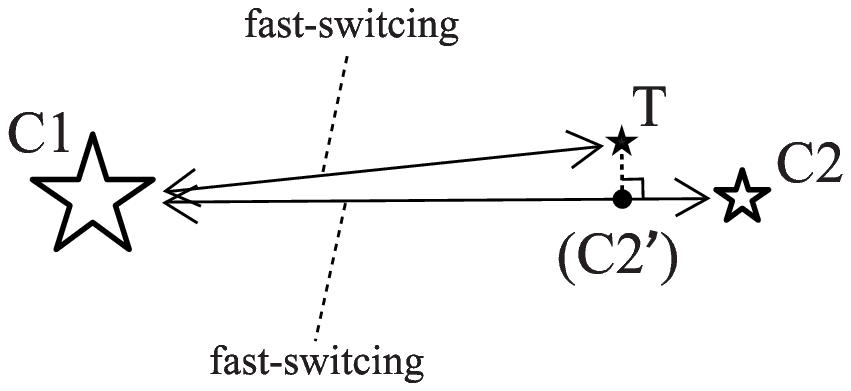}
  \end{center}
  \caption{An example of source configuration on the sky plane, and scheduling.  Only a primary calibrator C1 is fringe-detectable.  C2 is a secondary calibrator, a weak source.  C2$^\prime$ is a hypothetical calibrator, on blank sky at the tangential point on C1--C2 line.
}\label{fig:sourceconfig}
\end{figure}

The BPR method schedules three sources: a target `T', a primary calibrator `C1', and a secondary calibrator `C2'.  C1 is bright enough to detect in half of a coherence time, typically a few minutes or less, depending on observing frequency and weather condition \citep{Ulvestad1998}.  We assume here C2 provides correlated flux densities insufficient for fringe detection in half of the coherence time but sufficient for detection by phase-referencing of several tens of minutes: it is $\sim$10~mJy at centimeter bands for typical VLBI arrays.  On the basis of statistical discussion on the surface-number density of radio sources in the sky (e.g., \cite{Fomalont_etal.1991}), C1 may be located relatively far from T, typically $\sim$2~deg separation, while C2 can be found at the position significantly closer to T, possibly $<$0.5~deg separation.  An example of source configuration is shown in Fig.~\ref{fig:sourceconfig}.

Both T and C2 should be observed using phase-referencing because of their weakness.  An example of schedule for the BPR is as follows:
\begin{eqnarray}
& &\mathrm{\ldots-C1-C2-C1-C2-C1-C2-C1-C2}\nonumber \\ 
& &\ \ \ \ \mathrm{-C1\ -T-C1\ \ -T-C1\ -T-C1\ \ -T}\nonumber \\
& &\ \ \ \ \mathrm{-C1-C2-C1-C2-C1-C2-C1-C2}\nonumber \\
& &\ \ \ \ \mathrm{-C1\ -T-C1\ \ -T-C1\ -T-C1\ \ -T-\ldots}\label{equation:schedule}
\end{eqnarray}
It consists of sets of fast-switching for two pairs: C2--C1 and T--C1.  The schedule intends to remove rapid phase fluctuations by fast-switching, and to change the reference point C1 into C2 by pair-swapping in order to reduce the separation angle from T.  The scans for the C2--C1 pair must be allocated every less than 1~hour, so that the long-term phase drifts can be tracked.  The number of iterations of fast-switching for the C2--C1 pair should be set to secure the detection of C2 with a signal-to-noise ratio of more than 5, for successful self-calibration in phase domain.

\subsection{Observation equations}
With such an observing schedule, an observer will obtain raw visibility data for the target and two calibrators.  After determination of amplitude-gain, delay and delay-rate solutions using C1, all we have to consider is phase terms.  Observed, raw visibility phases, $\phi_\mathrm{OBS}$, involve various phase terms.
\begin{eqnarray}
\phi_\mathrm{OBS}^\mathrm{C1} &=& \phi_\mathrm{stru}^\mathrm{C1} + \phi_\mathrm{pos}^\mathrm{C1} + \phi_\mathrm{inst}^\mathrm{C1} + \phi_\mathrm{geo}^\mathrm{C1} + \phi_\mathrm{atmo}^\mathrm{C1} + \phi_\mathrm{rapid}^\mathrm{C1}\\
\phi_\mathrm{OBS}^\mathrm{C2} &=& \phi_\mathrm{stru}^\mathrm{C2} + \phi_\mathrm{pos}^\mathrm{C2} + \phi_\mathrm{inst}^\mathrm{C2} + \phi_\mathrm{geo}^\mathrm{C2} + \phi_\mathrm{atmo}^\mathrm{C2} + \phi_\mathrm{rapid}^\mathrm{C2}\\
\phi_\mathrm{OBS}^\mathrm{T} &=& \phi_\mathrm{stru}^\mathrm{T} + \phi_\mathrm{pos}^\mathrm{T} + \phi_\mathrm{inst}^\mathrm{T} + \phi_\mathrm{geo}^\mathrm{T} + \phi_\mathrm{atmo}^\mathrm{T} + \phi_\mathrm{rapid}^\mathrm{T},
\label{obs-phi}
\end{eqnarray}
where, $\phi^\mathrm{C1}$, $\phi^\mathrm{C2}$, and $\phi^\mathrm{T}$ are phase terms for C1, C2 and T, respectively.  
$\phi_\mathrm{stru}$ is a phase term originating in source structure; 
$\phi_\mathrm{pos}$ is a positional-phase delay relative to a phase-tracking center; 
$\phi_\mathrm{inst}$ is an instrumental-phase delay; 
$\phi_\mathrm{geo}$ is a geometric-phase delay error; 
$\phi_\mathrm{atmo}$ is tropospheric/ionospheric-phase delay error; 
$\phi_\mathrm{rapid}$ is a rapidly variable phase due to water vapor flowing at low altitude.  We ignore thermal phase noise and calibration errors in the present paper.  
We deal with the phase terms of source position and structure separately, in order to show an astrometric term and its error expressly at the last equation.    Time scales of the change of these phase-error components are usually more than several tens of minutes, except for $\phi_\mathrm{rapid}$ in which the time scale is typically a few minutes or less.

\begin{table*}
\caption{Array configurations for our observations.}\label{tableantenna}
\begin{center}
\begin{tabular}{lll} 
\hline
Epoch & Date & Telescopes \\\hline
1st & 2005 Sep 25 08:00-12:30UT & VERA$\times4$ (Mizsawa20m, Ogasawara20m, Iriki20m, Ishigaki20m) \\
2nd & 2005 Oct 23 06:00-12:00UT & VERA$\times4$, Kashima34m, Tsukuba32m, Usuda64m, Yamaguchi32m \\\hline
\end{tabular}
\end{center}
\end{table*}

\subsection{Calibrations of bigradient phase referencing}\label{section:calibration}  

Data of C1 have high signal-to-noise ratios enough to do self-calibration.  Hence, a sufficiently feasible source structure model can be obtained:
\begin{equation}
\phi_\mathrm{OBS}^\mathrm{C1} = \Phi_\mathrm{stru}^\mathrm{C1} + \Phi_\mathrm{SN1}, \label{eq:selfcalC1}
\end{equation}
where
\begin{equation}
\Phi_\mathrm{SN1}=\phi_\mathrm{pos}^\mathrm{C1} + \phi_\mathrm{inst}^\mathrm{C1} + \phi_\mathrm{geo}^\mathrm{C1} + \phi_\mathrm{atmo}^\mathrm{C1} + \phi_\mathrm{rapid}^\mathrm{C1}.
\end{equation}
An antenna-based solution table, provided from the self-calibration, includes all the terms other than the source structure.  This is a `solution table 1,' $\Phi_\mathrm{SN1}$.  We express a determined term in its capital, in the present paper.

With this solution table, we can do phase correction to both C2 and T, so that their phases are free from rapid phase fluctuations due to water vapor.  The calibration on C2 is as follows, 
\begin{eqnarray}
\phi_\mathrm{OBS}^\mathrm{C2} &-& \Phi_\mathrm{SN1}\nonumber \\
			      &=& \phi_\mathrm{stru}^\mathrm{C2} + \phi_\mathrm{pos}^\mathrm{C2} + \phi_\mathrm{inst}^\mathrm{C2} + \phi_\mathrm{geo}^\mathrm{C2} + \phi_\mathrm{atmo}^\mathrm{C2} + \phi_\mathrm{rapid}^\mathrm{C2}\nonumber \\
        & &\ \ \ \ \ \ - (\phi_\mathrm{pos}^\mathrm{C1} + \phi_\mathrm{inst}^\mathrm{C1} + \phi_\mathrm{geo}^\mathrm{C1} + \phi_\mathrm{atmo}^\mathrm{C1} + \phi_\mathrm{rapid}^\mathrm{C1})\nonumber \\
&=& \phi_\mathrm{stru}^\mathrm{C2} + \Delta\phi_\mathrm{pos}^\mathrm{C2-C1} + \Delta\phi_\mathrm{inst}^\mathrm{C2-C1} + \Delta\phi_\mathrm{geo}^\mathrm{C2-C1}\nonumber \\
& &\ \ \ \ \ \ \ + \Delta\phi_\mathrm{atmo}^\mathrm{C2-C1} + \Delta\phi_\mathrm{rapid}^\mathrm{C2-C1},\label{eq:SN1applied}
\end{eqnarray}
where $\Delta\phi_i^\mathrm{C2-C1} \equiv \phi_i^\mathrm{C2}-\phi_i^\mathrm{C1}$, differential phase terms.  Presumably, $\Delta\phi_\mathrm{rapid}^\mathrm{C2-C1} \approx 0$, because its random fluctuation is not responsible for a systematic residual, so should be averaged out.  Also $\Delta\phi_\mathrm{inst}^\mathrm{C2-C1} \approx 0$, because an identical receiving system is used.  The calibration on T can be done in the same manner; this is a normal phase-referencing, which has been commonly used \citep{Beasley&Conway1995}.

The other differential phase terms may not be zero.  They are responsible for long-term phase drifts, which may cause some apparent position shift and the degradation of image dynamic range.  As we mentioned above, the time scales of change of these errors are usually more than several tens of minutes; we can integrate the data coherently in a period of less than their time scales.  When signal-to-noise ratios sufficient to do self-calibration are available on C2 by this integration, its source structure model can be determined.  Then, equation~(\ref{eq:SN1applied}) becomes
\begin{equation}
\phi_\mathrm{OBS}^\mathrm{C2} - \Phi_\mathrm{SN1} \approx \Phi_\mathrm{stru}^\mathrm{C2} + \Phi_\mathrm{SN2},
\label{eq:selfcalC2}
\end{equation}
where
\begin{equation}
\Phi_\mathrm{SN2} \equiv \Delta\phi_\mathrm{pos}^\mathrm{C2-C1} + \Delta\phi_\mathrm{geo}^\mathrm{C2-C1} + \Delta\phi_\mathrm{atmo}^\mathrm{C2-C1}.\label{eq:solutiontable2}
\end{equation}
Thus, we obtain a `solution table 2,' $\Phi_\mathrm{SN2}$.  We recommend that the first self-calibration on C2 is done using a tentative point source model at the phase-tracking center, in order to include $\phi_\mathrm{pos}^\mathrm{C2}$ into the solution table~2 for astrometry.

Now we apply both the solution table~1 and 2 to the target data;
\begin{eqnarray}
\phi_\mathrm{OBS}^\mathrm{T} &-& (\Phi_\mathrm{SN1} + \Phi_\mathrm{SN2})\nonumber\\
	&=& \phi_\mathrm{stru}^\mathrm{T} + \phi_\mathrm{pos}^\mathrm{T} + \phi_\mathrm{inst}^\mathrm{T} + \phi_\mathrm{geo}^\mathrm{T} + \phi_\mathrm{atmo}^\mathrm{T} + \phi_\mathrm{rapid}^\mathrm{T}\nonumber\\
	& & \ \ \ \ \ \ - [ \phi_\mathrm{pos}^\mathrm{C1} + \phi_\mathrm{inst}^\mathrm{C1} + \phi_\mathrm{geo}^\mathrm{C1} + \phi_\mathrm{atmo}^\mathrm{C1} + \phi_\mathrm{rapid}^\mathrm{C1}\nonumber\\
        & & \ \ \ \ \ \ + \Delta\phi_\mathrm{pos}^\mathrm{C2-C1} + \Delta\phi_\mathrm{geo}^\mathrm{C2-C1} + \Delta\phi_\mathrm{atmo}^\mathrm{C2-C1} ]\nonumber\\
	&=& \phi_\mathrm{stru}^\mathrm{T} + \Delta\phi_\mathrm{pos}^\mathrm{T-C2} + \Delta\phi_\mathrm{inst}^\mathrm{T-C1} + \Delta\phi_\mathrm{rapid}^\mathrm{T-C1}\nonumber\\
 	& & \ \ \ \ \ \ \ \ \ \ \ \ \ \ \ \ \ \ \ \ \ + [ \Delta\phi_\mathrm{geo}^\mathrm{T-C2} + \Delta\phi_\mathrm{atmo}^\mathrm{T-C2} ].\label{eq:BPR}
\end{eqnarray}
This equation is very similar to equation~(\ref{eq:SN1applied}), but this includes the T--C2 pair.  Again $\Delta\phi_\mathrm{inst}^\mathrm{T-C1} \approx 0$ and $\Delta\phi_\mathrm{rapid}^\mathrm{T-C1} \approx 0$.  The process in this equation has replaced C1 with C2 as a reference point, where array's focus is strictly optimized.  In other words, we can make a substantially phase-referencing for the T--C2 pair without fast-switching between them.  A fringe-undetectable source becomes useful as a reference calibrator.  The separation angle of the T--C2 pair is smaller than that of the T--C1 pair, so $\Delta\phi_\mathrm{geo}^\mathrm{T-C2} < \Delta\phi_\mathrm{geo}^\mathrm{T-C1}$ and $\Delta\phi_\mathrm{atmo}^\mathrm{T-C2} < \Delta\phi_\mathrm{atmo}^\mathrm{T-C1}$.  This means that the differential excess path becomes smaller, and we would obtain a better image quality in comparison with that of the normal phase-referenced image.  This is the bigradient phase-referencing.  The astrometric measurement of T can be done relative to C2, because of $\Delta\phi_\mathrm{pos}^\mathrm{T-C2}$, with the uncertainty of $\Delta\phi_\mathrm{geo}^\mathrm{T-C2} + \Delta\phi_\mathrm{atmo}^\mathrm{T-C2}$, as shown equation~(\ref{eq:BPR}).

In most cases, observers have to be careful about $\phi_\mathrm{pos}^\mathrm{C2}$ in the stage of self-calibration on C2 [equation~(\ref{eq:selfcalC2}) and (\ref{eq:solutiontable2})].  The mas-scale position of C2 may be unknown when scheduling, because such a weak source is not cataloged into the International Celestial Reference Frame (ICRF; \cite{Ma_etal.1998}).  In the phase-referenced image of C2 [equation~(\ref{eq:SN1applied})], an emission peak may initially appear shifted by several tens of mas, even if the phase-tracking center was set based on a position measured with the Very Large Array~(VLA) A-array configuration, etc.  This leads a too large $\Phi_\mathrm{SN2}$ to make phase-connections beyond 2$\pi$ ambiguity.  Also, the requirement of position accuracy of C2 is the same as that of C1, because C2 has become a reference calibrator as a substitute for C1 in the BPR.  The large position error of a reference calibrator will degrade the dynamic range of target image (section~17.3.5 of \cite{Beasley&Conway1995}).  We recommend correcting the phase-tracking center to an accurate position by post processing with the tasks CLCOR or UVFIX in the AIPS, before the self-calibration on C2.

\subsection{A more-sophisticated calibration}\label{section:moresophis}
The solution table~2, $\Phi_\mathrm{SN2}$, has tracked the long-term phase drift between the C2--C1 pair, shown as equation~(\ref{eq:solutiontable2}).  However, we hope ultimately to know that between the T--C1 pair.  The method of \citet{Fomalont&Kogan2005} achieves it using more than one calibrator around a target, in the assumption that long-term phase variations are linear over the region of the sources \citep{Fomalont&Kopeikin2003}.  We also apply the same assumption to an additional calibration, which is more-sophisticated than the BPR.  Array's focus has moved to C2 from C1 by the BPR (section~\ref{section:calibration}).  This means the solution table~2 has the ability to shift the focus.  We can make a new focus at any point on the line of C2--C1 as far as the linearity is valid.  In the more-sophisticated calibration, we establish the hypothetical calibrator C2$^\prime$, on blank sky at the tangential point on the C1--C2 line from T, by scaling 
$\Phi_\mathrm{SN2}$ with a factor $r$, where $r \equiv (\theta^\mathrm{C1-C2^\prime} / \theta^\mathrm{C1-C2})$, $\theta^\mathrm{C1-C2^\prime}$ and $\theta^\mathrm{C1-C2}$ are separation angles of C1--C2$^\prime$ and C1--C2, respectively (Fig.~\ref{fig:sourceconfig}).  If both the solution table~1 and this modified solution table~2 are applied to the target data, 
\begin{eqnarray}
\phi_\mathrm{OBS}^\mathrm{T} &-& (\Phi_\mathrm{SN1} + r \times \Phi_\mathrm{SN2})\nonumber\\
	&=& \phi_\mathrm{stru}^\mathrm{T} + \phi_\mathrm{pos}^\mathrm{T} + \phi_\mathrm{geo}^\mathrm{T} + \phi_\mathrm{atmo}^\mathrm{T} + \phi_\mathrm{inst}^\mathrm{T} + \phi_\mathrm{rapid}^\mathrm{T}\nonumber\\
	& & \ \ \ \ -\ [ ( \phi_\mathrm{pos}^\mathrm{C1} + \phi_\mathrm{geo}^\mathrm{C1} + \phi_\mathrm{atmo}^\mathrm{C1} + \phi_\mathrm{inst}^\mathrm{C1} + \phi_\mathrm{rapid}^\mathrm{C1} )\nonumber\\
        & & +\ r \times ( \Delta\phi_\mathrm{pos}^\mathrm{C2-C1} + \Delta\phi_\mathrm{geo}^\mathrm{C2-C1} + \Delta\phi_\mathrm{atmo}^\mathrm{C2-C1}) ]\nonumber\\
	&=& \phi_\mathrm{stru}^\mathrm{T}+ \Delta\phi_\mathrm{pos}^\mathrm{T-C1} - r \times \Delta\phi_\mathrm{pos}^\mathrm{C2-C1}\nonumber\\
	& & + \Delta\phi_\mathrm{inst}^\mathrm{T-C1} + \Delta\phi_\mathrm{rapid}^\mathrm{T-C1}\nonumber\\
	& & + [\ (\Delta\phi_\mathrm{geo}^\mathrm{T-C1} - r \times \Delta\phi_\mathrm{geo}^\mathrm{C2-C1})\nonumber\\
	& & +\ \ (\Delta\phi_\mathrm{atmo}^\mathrm{T-C1} - r \times \Delta\phi_\mathrm{atmo}^\mathrm{C2-C1})\ ].\label{eq:moresophi0}  
\end{eqnarray}
Again $\Delta\phi_\mathrm{inst}^\mathrm{T-C1} \approx 0$ and $\Delta\phi_\mathrm{rapid}^\mathrm{T-C1} \approx 0$.  Because of the linearity, the terms $r \times \Delta\phi_\mathrm{geo}^\mathrm{C2-C1}$ and $r \times \Delta\phi_\mathrm{atmo}^\mathrm{C2-C1}$ virtually correspond to differential geometric and atmospheric phase-delays between C2$^\prime$ and C1: $r \times \Delta\phi_\mathrm{geo}^\mathrm{C2-C1} = \Delta\phi_\mathrm{geo}^\mathrm{C2^\prime-C1}$ and $r \times \Delta\phi_\mathrm{atmo}^\mathrm{C2-C1} = \Delta\phi_\mathrm{atmo}^\mathrm{C2^\prime-C1}$.  Finally, equation~(\ref{eq:moresophi0}) becomes 
\begin{eqnarray}
\phi_\mathrm{OBS}^\mathrm{T} &-& (\Phi_\mathrm{SN1} + r \times \Phi_\mathrm{SN2})\nonumber\\
	&\approx& \phi_\mathrm{stru}^\mathrm{T}+ \Delta\phi_\mathrm{pos}^\mathrm{T-C1} - r \times \Delta\phi_\mathrm{pos}^\mathrm{C2-C1}\nonumber\\
	& & + [\ \Delta\phi_\mathrm{geo}^\mathrm{T-C2^\prime} + \Delta\phi_\mathrm{atmo}^\mathrm{T-C2^\prime}\ ].\label{eq:moresophi}  
\end{eqnarray}
This is substantially a phase-referencing for the T--C2$^\prime$ pair, although there is no fast-switching between them.  The modified phase solutions of self-calibration have made array's focus at the sky toward C2$^\prime$, much closer to the target.

The astrometric term in equation~(\ref{eq:moresophi}) is $\Delta\phi_\mathrm{pos}^\mathrm{T-C1}$.  The term $r \times \Delta\phi_\mathrm{pos}^\mathrm{C2-C1}$ consists of only positional terms, so it makes only a slight position shift without degrading an image dynamic range.  Even with ICRF calibrators, their absolute-position uncertainties ($\sim0.3$~mas, \cite{Ma_etal.1998,Fey_etal.2004}) may be responsible for the term.  Since the differences between the real positions and phase-tracking center are unpredictable, $r \times \Delta\phi_\mathrm{pos}^\mathrm{C2-C1}$ is also unpredictable.  However, this will practically not contribute to astrometric errors in observations of relative astrometry, because $r \times \Delta\phi_\mathrm{pos}^\mathrm{C2-C1}$ is duplicated every epoch if the relative positions between the emission peaks of calibrators and the phase-tracking centers are not changed any epochs.  Thus, we can make relative astrometry for measuring proper motion in series of monitoring observations.  Please note that observer must use identical phase-tracking centers every epoch and the first self-calibrations using the structure models at the phase-tracking centers.

The terms of $[\Delta\phi_\mathrm{geo}^\mathrm{T-C2^\prime} + \Delta\phi_\mathrm{atmo}^\mathrm{T-C2^\prime}]$ are expected to be zero, if the position of C2$^\prime$ is coincident with that of T, i.e., all the three sources align exactly on a straight line.  If the sources are misaligned, these terms become a phase error, depending on the separation angle between C2$^\prime$ and T.  Because the BPR makes a lot of weaker calibrators usable, observers will have a desirable allocation of sources in the sky much easier.  Possible solutions to reduce this remaining error of the misaligned case are discussed in Section~\ref{section:implication}.

\subsection{Calibration errors}\label{section:calibrationerror}
In addition to the errors theoretically expected in the method, there are various calibration errors.  These kinds of errors are usually not so large compared to the long-term phase drifts, but still sufficient to affect the quality of a phase-referenced image.  Observers must be careful to address them as much as possible, when pursuing accurate astrometry.  

(1) The purity of the solution tables is related to the quality of self-calibration in equations~(\ref{eq:selfcalC1}) and (\ref{eq:selfcalC2}).  Very complex source structures or low signal-to-noise ratios make it difficult to separate $\Phi_\mathrm{stru}$ from the other phase terms.  This calibration error would be significant, because the BPR involves self-calibration twice, and because one of two calibrators is always a weak source.  An antenna-based phase solution with a signal-to-noise ratio of~3 corresponds to an accuracy of $\sim21$~deg theoretically \citep{Thompson_etal.2001}, which would be as large as systematic phase components to be corrected.  Hence, we recommend the use of C2 with an signal-to-noise ratio of more than~5.  Note that phase signals are usually weaker on longer baselines.  

(2) The rapid phase fluctuations with the same as or less than the period of fast-switching cannot be removed, although we assumed $\Delta\phi_\mathrm{rapid}^\mathrm{C2-C1} \approx 0$ and $\Delta\phi_\mathrm{rapid}^\mathrm{T-C1} \approx 0$.  The remaining fluctuations tend to be random and averaged out, but are responsible for coherence loss.  This situation also arises in normal phase-referencing of fast-switching.  A dual-beam observation, as the VERA, is presumably the most effective ways against such a random phase fluctuation.

(3) The longer term changes of $\phi_\mathrm{pos}$, $\phi_\mathrm{atmo}$, and $\phi_\mathrm{geo}$ have time scales of more than several tens of minutes.  The pair-swapping tracks them, but cannot monitor their fluctuations with less time scales due to absence of fast-switching between T and C2.  The fluctuations affect not only phase measurements in C2 scans but also calibration parameters that will be made for T by interpolation between the C2 scans.  We recommend that the pair-swapping period is about 20~minutes, which provides some redundancy in the time scale of long-term phase drifts, when accurate astrometry is required.  The fluctuations will be averaged out, although some cohelence loss is not avoidable.  

(4) The structure change of calibrators results in a position shift of the target; astrometric errors of 100~microarcseconds over observations longer than several months are possible.  Although this is not a kind of calibration errors, 
it is not avoidable for observers.  The intrinsic position change on a calibrator will bring the same position shift on the target without degrading its image dynamic range.  It is necessary to use calibrators with stable structures, preferably point sources; do not use super-luminul quasars even if high signal-to-noise ratios are expected.

\section{Test observations}\label{section:testobs}
We present demonstrative VLBI observations by the BPR method in this section.  Although this method aims to detect a weak target using a weak calibrator and a strong calibrator, for easy inspection of visibility-phase correction, here we selected a target and two calibrators from strong ICRF radio sources, as follows:
\begin{itemize}
\item 3C~345 as a target `T'
\item NRAO~512 as a secondary calibrator `C2'
\item DA~426 as a primary calibrator `C1.'
\end{itemize}
The relative locations of these sources are similar to Fig.~\ref{fig:sourceconfig}, but align more straightly on east--west direction on the sky plane.  C1 and T, a primary pair, are separated by 2.09~deg; C1 and C2, a secondary pair, are separated by 2.57~deg; C2 and T are separated by 0.48~deg.  T and C2$^\prime$ are separated by only 2.1~arcmin. 


\subsection{Observations}
VLBI imaging observations with the BPR method were carried out at two epochs with the Japanese VLBI network~(JVN), a newly-established VLBI network with $\sim50$-2300~km baselines across the Japanese islands (\authorcite{Fujisawa_etal.inprep} in~prep).  This array consists of ten antennas, including four radio telescopes of the VLBI Exploration of Radio Astrometry project (VERA; \cite{Kobayashi_etal.2003}).  The telescope participants for the observations are listed in Table~\ref{tableantenna}.  Right-circular polarization was received at X-band.  Two frequency bands, 8400-8416~MHz (IF1) and 8432-8448~MHz (IF2), were selected.  The VSOP-terminal system was used as a digital back-end; digitized data in 2-bit quantization were recorded onto magnetic tapes at a data rate of 128~Mbps.  Correlation processing was performed with the Mitaka FX correlator \citep{Shibata_etal.1998} at the National Astronomical Observatory Japan.

The switching-cycle period of phase-referencing was 5 minutes.  A set of switching cycles for the primary pair~(T--C1) spent about 20 minutes; a set for the secondary pair~(C2--C1) also spent about 20 minutes.  The two sets were scheduled alternately, as mentioned in Section~\ref{section:arrange&schedule}.  Thus, a pair-swapping cycle period was about 40~minutes.

\subsection{Data reduction}
Data reduction procedures were performed for three data sets: 1st-epoch data~(VERA antennas), 2nd-epoch data from only the VERA antennas, and 2nd-epoch data from all the antennas (Table~\ref{tableantenna}).  The correlated data were reduced using the AIPS.  After initial data inspection and flagging, fringe-fitting was performed to 3C~345 and NRAO~512.  Note that in this stage we did not intend to use the two sources as T or C2 but for bandpass calibration and amplitude-gain calibration, respectively.  Since NRAO~512 is the most ideal point source up to our longest baseline~($\sim50$~M$\lambda$), self-calibration in terms of amplitude provided us the solution of amplitude-gain variation relative to each antenna with the accuracy to about 1\%.  We have done absolute-amplitude scaling using NRAO~512 with 0.89$\pm$0.09~Jy, a single-dish measurement with the Yamaguchi 32-m radio telescope \citep{Fujisawa_etal.2002} at 8.38~GHz on 2005 Nobember~12.  After these processes, we deleted the solutions of fringe-fitting for 3C~345 and NRAO~512.  Data obtained at elevations of less than $\sim25$~deg were flagged out.  We utilized no a-priori gain parameter in both amplitude and phase except for the absolute flux scaling.

Next, we performed fringe-fitting for C1~(DA~426), then established its source structure model in the DIFMAP software \citep{Shepherd1997} using deconvolution and self-calibration algorithms iteratively.  With the CALIB task in the AIPS, the phase solution of self-calibration was obtained using the source structure model: we obtained a `solution table 1' (section~\ref{section:calibration}).  We applied this table to the data of both T and C2.  At this stage, we can obtain normal phase-referenced images.

\begin{figure*}[htb]
  \begin{center}
    \FigureFile(160mm,160mm){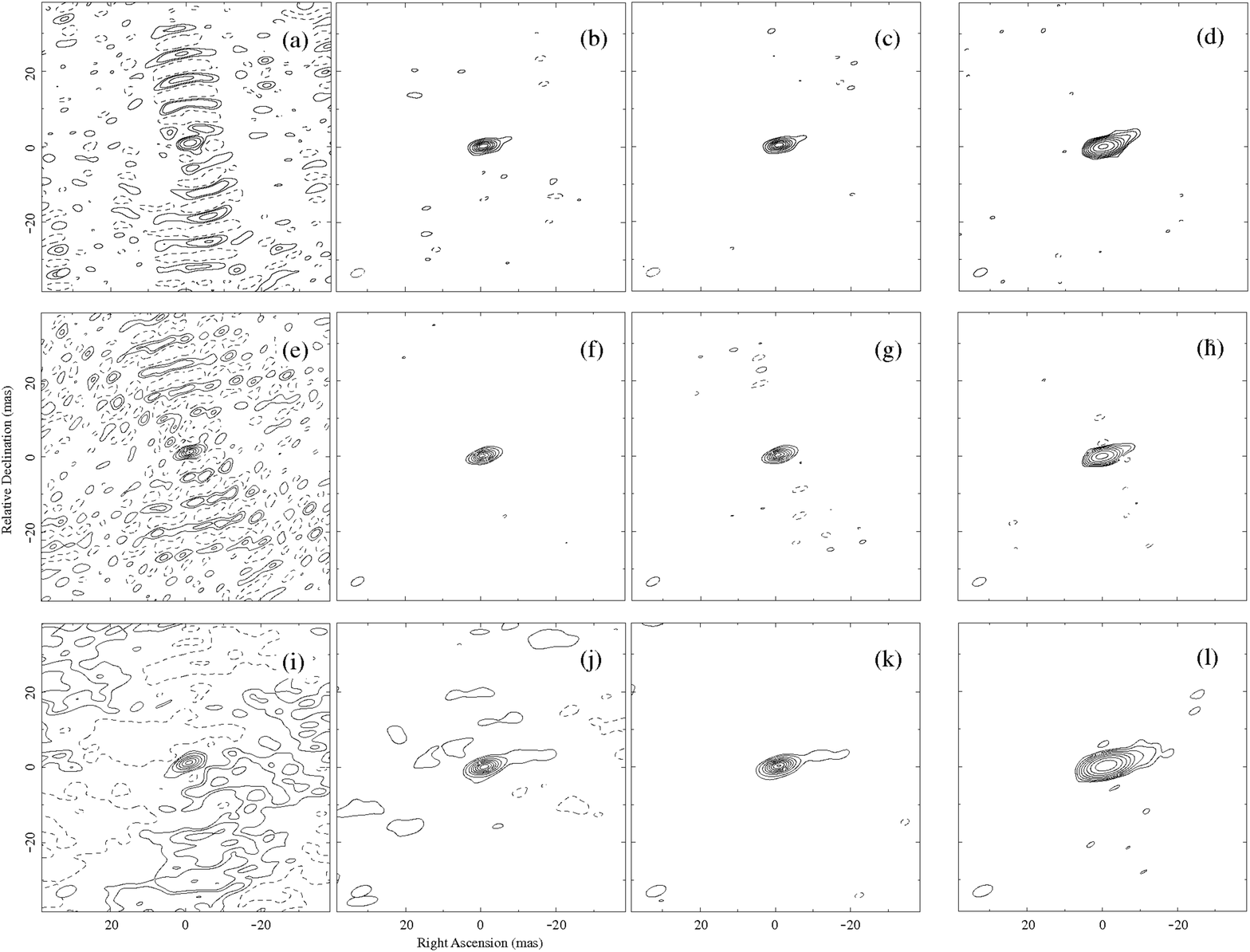}
  \end{center}
  \caption{Contour maps of target 3C~345.  {\bf (upper panels)} Images at the first epoch with four telescopes of the VERA.  (a)~normal phase-referenced image.  (b)~Bigradient phase-referenced image.  (c)~Image made by more-sophisticated calibration.  Contours of (a)-(c) start from $3\sigma_\mathrm{c}$, where $\sigma_\mathrm{c}$ is RMS of image noise on the image~(c) (see Table~\ref{table:result}).  (d)~Self-calibrated image.  Contours of (d) start from $3\sigma_\mathrm{d}$, where $\sigma_\mathrm{d}$ is RMS of image noise on the image~(d).  Synthesized beams, shown at the lower-left of each image, are of $3.75 \times 2.02$~mas at a position angle of $-$69.3~deg.  
{\bf (middle panels)} Images at the second epoch with only four telescopes of the VERA.  (e)~normal phase-referenced image.  (f)~Bigradient phase-referenced image.  (g)~Image made by more-sophisticated calibration.  Contours of (e)-(g) start from $3\sigma_\mathrm{g}$.  (h)~Self-calibrated image.  Contours of (h) start from $3\sigma_\mathrm{h}$.  Synthesized beams are of $2.00 \times 3.52$~mas at a position angle of $-$67.3~deg.  
{\bf (lower panels)} Images at the second epoch with all telescopes.  (i)~normal phase-referenced image.  (j)~Bigradient phase-referenced image.  (k)~Image made by more-sophisticated calibration.  Contours of (i)-(k) start from $3\sigma_\mathrm{k}$.  (l)~Self-calibrated image.  Contours of (l) start from $3\sigma_\mathrm{l}$.  Synthesized beams are of $2.72 \times 5.54$~mas at a position angle of $-$70.6~deg.  
Contour levels for phase-referenced images are $3\sigma \times(-1$, 1, 2, 4, 6, 8, 10, 12, 14, 16); for self-calibrated images are $3\sigma \times(-1$, 1, 2, 4, 8, 16 ,32 ,64 ,128 ,256).  Negative contours are described in dashed-curves.}\label{fig:image} 
\end{figure*}

The phase-referenced visibilities of C2~(NRAO~512) were coherent in at least several tens of minutes~(described in Section~\ref{section:longtermphasedrift}), as we have assumed.  C2 was supposed to have sufficient signal-to-noise ratios to perform self-calibration in phase when integrated for several tens of minutes, even if with down to $\sim$10~mJy.  We have successfully obtained a `solution table 2' by self-calibration for NRAO~512 in a solution interval of 40~minutes.  The target image, which were synthesized from visibilities corrected with both the calibration table 1 and 2, are shown in Fig.~\ref{fig:image}~(b), (f) and~(j); these are the images provided by the BPR method.

Since the three sources quasi-perfectly align on a straight line on the sky.  This is an ideal situation for the more-sophisticated calibration (Section~\ref{section:moresophis}).  The phase solutions in the solution table~2 was scaled down to $r$~$(= [\theta^\mathrm{C1-C2^\prime} / \theta^\mathrm{C1-C2}]$) times, and then the modified solution table was applied to the target data together with the solution table~1.  The resultant images are shown in Fig.~\ref{fig:image}~(c), (g), and~(k).  For comparison, images corrected by self-calibration both in amplitude and phase are also shown in Fig.~\ref{fig:image}~(d), (h), and~(l).  

The number of usable visibilities in the data set of 2nd-epoch from only the VERA antennas were $\sim 30$\% less than the 1st-epoch ones, because the significant fraction of C1's scans were flagged out in fringe-fitting.  With such a small number of visibilities and a poor UV-coverage, the resultant images could not bear comparison with the other data sets.

\subsection{Results}

\subsubsection{Image quality}
Images by the BPR appeared in Fig.~\ref{fig:image} are dramatically improved compared to those by normal phase-referencing: image dynamic ranges have increased about 3.1-3.8 times (Table~\ref{table:result}).  The jet structure of 3C~345 is seen in the improved images.  This is the fruit of using a closer calibrator adopted in the BPR.  The normal phase-referencing was from the T--C1 pair, separated by 2.09~deg, while the BPR was practically from the T--C2 pair, separated by only 0.48~deg, a 4.3-times smaller separation.  Better images have been obtained by the more-sophisticated calibration.  This is the fruit of optimization of the solution table~2 for the sky point of C2$^\prime$, separated from T by only 2.1~arcmin, a 61-times smaller separation than that of T--C1.  In the VERA image at the 2nd epoch, the more-sophisticated calibration appears not to work well in terms of a image-noise level.  Because of small number of effective visibility of this data set, measurements for this data set might be some inadequate.

\begin{table*}[hbt]
\caption{Phase-referencing qualities in images of 3C~345.}\label{table:result}
\begin{center}
\begin{tabular}{lllcccccc} \hline
Epoch & Ant. & Method & $I^\mathrm{p}$ & $\sigma_\mathrm{img}$ & $I^\mathrm{p}/\sigma_\mathrm{img}$ & $\sigma_\phi$ & $\Delta\alpha$ & $\Delta\delta$ \\
 &  &  & (Jy beam$^{-1}$) & (mJy beam$^{-1}$) &  & (deg) & (mas) & (mas) \\
\multicolumn{1}{c}{(1)} & \multicolumn{1}{c}{(2)} & \multicolumn{1}{c}{(3)} & (4) & (5) & (6) & (7) & (8) & (9) \\\hline
1st & VERA & PR & 3.49  & 291 & \ 12 & 65 & $-1.103 \pm 0.118$ & $+0.925 \pm 0.059$ \\
 &  & BPR & 4.03  & 107 & \ 37 & 26 & $-0.956 \pm 0.054$ & $+0.238 \pm 0.025$ \\
 &  & BPR$+\alpha$ & 4.06  & \ \ \ \ 97.4 & \ 41 & 30 & $-0.999 \pm 0.050$ & $+0.374 \pm 0.023$ \\
 &  & self-cal. & 4.32  & \ \ \ \ \ \ \ 9.23 & 468 & 21 & \ldots & \ldots \\
2nd & VERA & PR & 3.62  & 290 & \ 12 & 69 & $-1.308 \pm 0.159$ & $+1.198 \pm 0.074$ \\
 &  & BPR & 3.90  & \ \ \ \ 88.7 & \ 44 & 21 & $-0.725 \pm 0.044$ & $+0.077 \pm 0.022$ \\
 &  & BPR$+\alpha$ & 3.93  & 100 & \ 39 & 28 & $-0.871 \pm 0.050$ & $+0.307 \pm 0.025$ \\
 &  & self-cal. & 4.10  & \ \ \ \ 15.8 & 259 & 19 & \ldots & \ldots \\
 & all & PR & 3.04  & 400 & \ \ 8 & 79 & $-0.926 \pm 0.313$ & $+1.243 \pm 0.175$ \\
 &  & BPR & 4.34  & 144 & \ 30 & 33 & $-1.001 \pm 0.083$ & $+0.126 \pm 0.043$ \\
 &  & BPR$+\alpha$ & 4.56  & \ \ \ \ 93.2 & \ 49 & 30 & $-0.967 \pm 0.055$ & $+0.320 \pm 0.028$ \\
 &  & self-cal. & 4.91  & \ \ \ \ \ \ \ 7.26 & 676 & 18 & \ldots & \ldots \\\hline
\end{tabular}
\end{center}
{\footnotesize Comments --- (1) observing epoch, (2) antennas used for mapping (3) calibration method. `PR' represents normal phase referencing. `BPR' represents bigradient phase referencing. `BPR$+\alpha$' represents more-sophisticated calibration.  `self-cal.' represents self-calibration. (4) peak intensity, (5) RMS of image noise, (6) image dynamic range, (7) RMS of visibility-phase scatter, (8) and (9) position shift from phase-tracking center of 3C~345. This was measured by two-dimensional Gaussian fitting using task JMFIT of AIPS.}
\end{table*}

\subsubsection{Long-term phase drifts}\label{section:longtermphasedrift}

\begin{figure*}[htb]
  \begin{center}
    \FigureFile(160mm,160mm){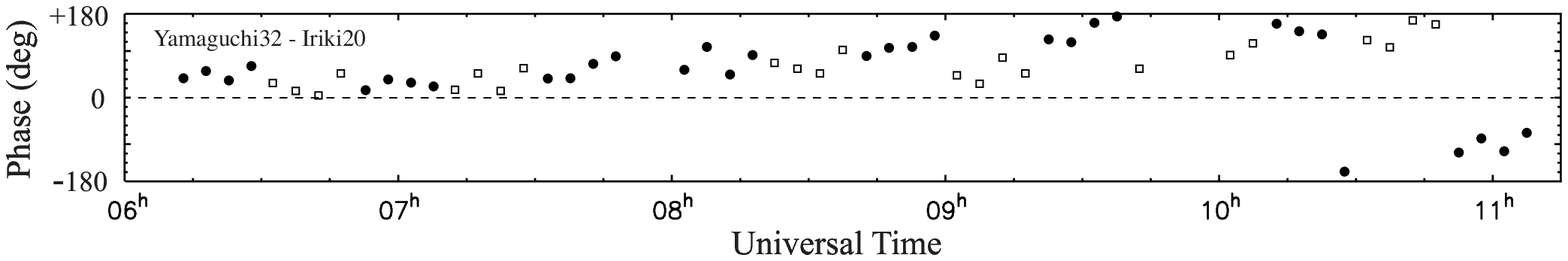}
  \end{center}
  \caption{Plot of visibility phases phase-referenced by primary calibrator DA~426 on Yamaguchi32m--Iriki20m baseline.  Open squares and filled circles represent phases of target 3C~345 and NRAO~512, respectively.  Visibilities were averaged for 5~mimutes in this plot.}\label{fig:vplot}
\end{figure*}

In the BPR method, phase-referenced visibilities of C2 are supposed to be coherent in at least several tens of minutes.  Additionally, the more-sophisticated calibration expects that the phase of phase-referenced C2 is a linear extrapolation of the phase-drift between T and C1.  We show an example of phase-referenced visibility phases both of 3C~345~(T) and NRAO~512~(C2) in Fig.~\ref{fig:vplot}.  A long-term phase-drift is seen.  The phase-drift of 3C~345 was undesirable for normal phase-referencing.  From DA~426's~(C1's) point of view, both the two sources are at about the same direction and separation.  The phases of NRAO~512 are supposed to be similar to the drift of 3C~345, and this expectation met the obserevation.  Consequently, we can remove a large part of the phase drift on 3C~345 by shifting array's focus to NRAO~512.

The phase drift of NRAO~512 became larger than that of 3C~345 especially later than 09h, where elevations of these sources became lower.  This deviation occurred because that the separation angle of NRAO~512--DA~426 was larger than that of 3C~345--DA~426.  Consequently, the data of 3C~345 will be over-corrected by the solution table~2 in the BPR.  The more-sophisticated calibration aims to correct such a difference by optimizing the solution table~2 toward the hypothetical calibrator C2$^\prime$.  The RMS of calibrated phases are listed in column~(7) of Table~\ref{table:result}.  Steady improvements were seen in the images from the 2nd epoch data from all the antennas.  However, slightly increases of RMS were seen in VERA data.  It may be a coincidental effect of the intrinsic structure of 3C~345; the VERA array is sensitive to source structure because of its longest baselines in the JVN array.  The improvements of the other parameters~(e.g., peak intensity) are the proof of efficacy of more-sophisticated calibration.

\subsubsection{Position shifts}\label{section:positionshift}

We evaluate here an astrometric capability from positional reproducibility between the two epochs.  Because of only a month, intrinsic changes of source structures were presumably negligible.  The measured position offsets from the phase-tracking center on the target images are listed in column~(8) and (9) of Table~\ref{table:result}.  The improvements of positional reproducibity by the BPR and more-sophisticated calibration are visualized in Fig.~\ref{fig:positionshift}.  The position differences between the 1st epoch and 2nd epoch of the target with all the available antennas were steadily reduced in our calibration steps: 360$\pm$240~micro-arcsecond~($\mu$as) between the normal phase-referenced images, 120$\pm$57~$\mu$as between the BPR images, and finally, 62$\pm$50~$\mu$as between the images by more-sophisticated calibration.    Globally, the emission peak of 3C~345 appeared at $\sim1$~mas northwest from the phase-tracking center.  These global position shifts include several origins, which will be discussed in Section~\ref{section:discussion_astrometry}.

\section{Discussions}\label{section:discussion}
\subsection{Phase-referencing capabilities}
The BPR substantially achieves phase-referencing between the C2--T pair without fast-switching between them.  In our tests, 3C~345 and NRAO~512 were separated by 0.48~deg, 4.3-times smaller than the separation angle of actual fast-switching between 3C~345 and DA~426.  A long-term phase drift from atmospheric/geometric errors, therefore, was expected to be 4.3-times smaller \citep{Beasley&Conway1995}.  However, resultant dynamic ranges of BPR images have been improved only 3.1-3.8 times better than those of normal phase-referenced images.  This discrepancy was probably caused by several calibration errors (Section~\ref{section:calibrationerror}).  The phase drifts seem to be successfully tracked with the sampling of C2 every 40~minutes.  However, this sampling frequency might be risky in some level, as seen in Fig.~\ref{fig:vplot} the phase of NRAO~512 and 3C~345 are systematically different at about 9.2h~UT.  It is advisable to pair-switch every about 20~minutes, or to apply an another observing schedule with more frequent C2 scans, such as $\mathrm{-C1-C2-C1-T-C1-C2-C1-T-C1-}$\ldots.  The observation case of indeed very weak C2, which is originally supposed in the BPR, will be reported in a future paper.  The BPR would give a better quality than that of the normal phase-referencing, even with a weak~($\sim$10~mJy) C2 with a signal-to-noise ratio of~5 in several scans using typical arrays at centimeter bands.

We briefly discuss the over-correction of the solution table~2 in the BPR method, by comparison of 2nd epoch images with all the antennas.  In the normal phase-referenced images both of 3C~345 and NRAO~512 (not shown), very similar patterns of distorted contours are seen: positive and negative contours reside mainly at lower-right and upper-left side corners in the images, respectively~[(i) in Fig.~\ref{fig:image}].  On the other hand, an opposite trend appears in the contours of the BPR image of 3C~345~[(j) in Fig.~\ref{fig:image}].  This is an effect of over-correction of solution table~2, which was designed for 2.57-degree separation of the C1--C2 pair rather than 2.09-degree separation of the C1--T pair.  The distortion trend almost disappears in the image by more-sophisticated calibration using the optimized table [(k) in Fig.~\ref{fig:image}].

\subsection{Astrometric capabilities}\label{section:discussion_astrometry}
The BPR and the more-sophisticated calibration support relative astrometry, if observers use identical phase-tracking centers every epochs (Section~\ref{section:principle}).  The consistency of the measured positions between our two epochs (Section~\ref{section:positionshift}) indicated that relative astrometry presumably worked well.  The position differences among the epochs are consistent with phase noises~(Table~\ref{table:result}).   

\begin{figure}[htp]
  \begin{center}
     \FigureFile(80mm,160mm){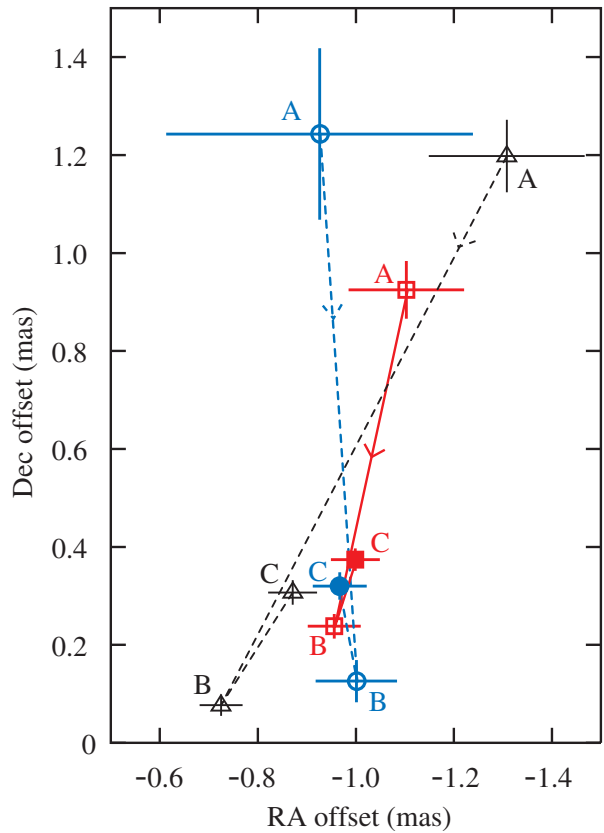}
  \end{center}
  \caption{Improvements of reproducibility of target's~(3C~345) position.  The origin (0,0) of this plot is the phase-tracking center of the target.  Labels A, B, and C show three different methods to measure positions.  A: normal phase-referencing, B: BPR, C: more-sophisticated calibration.  Three different symbols show different epochs and arrays.  Squares represent 1st-epoch measurements with VERA antennas.  Triangles represent 2nd-epoch measurements with VERA antennas.  Circles represent 2nd-epoch measurements with all antennas.  The most calibrated images show a position difference of only 62$\pm$50~$\mu$as between two epochs~(filled symbols).}\label{fig:positionshift}
\end{figure}

Dramatic revisions of the target position are seen particularly in declination (Fig.~\ref{fig:positionshift}).  This indicates that the long-term phase drifts were responsible for a large part of the declination offsets.  Geometric errors should be small in the VERA's four telescopes, Kashima34m, and Tsukuba32m, whose coordinates are strictly maintained by geodetic observations.  The correlator model in the Mitaka FX correlator seems to have somewhat prediction errors in zenith excess-path lengths at troposphere and/or ionosphere.  Note that a-priori calibration for such atmospheric errors is usually performed in the VERA project.

There remain significant offsets $\sim1$~mas west and $\sim300$~$\mu$as north in both the epochs, even with the more-sophisticated calibration.  Since such large global offsets could not be caused by any atmospheric/geometric errors in principle, an accumulation of intrinsic differences between actual emission's positions and phase-tracking centers of the target and two calibrators should be mainly responsible for the global offsets.  This kind of offsets cannot be avoided even with the ICRF radio sources and their cataloged positions, which have uncertainties of $\sim0.3$~mas \citep{Ma_etal.1998,Fey_etal.2004}, not only in the BPR and the more-sophisticated calibration but also in the normal phase-referencing.  DA~426~(C1) has mas-scale jets east-southeast; 3C~345~(T) has mas-scale jets west-northwest.  Hence, the global offsets pointing toward west-northwest may be reasonable.  

A position difference of 62$\pm$50~$\mu$as between the two epochs in the images by more-sophisticated calibration might originate in the slightly misaligned allocation of the three sources.  Strictly, the optimized solution table~2 reacted to the sky position of C2$^\prime$.  This position was separated by 2.1~arcmin from T.  The atmospheric and geometric errors from this separation could cause a $\sim10$-20~$\mu$as difference between the two epochs.  If the three sources perfectly aligned, the apparent positions of two epochs might coincide with each other within an error bar.

\subsection{Implications of the methods}\label{section:implication}
Although both the BPR and the more-sophisticated calibration aim to retaliate against long-term phase drifts by reducing a separation angle, they are based on slightly different tactics.  The BPR makes a weak source play a role of calibrator~C2.  Since self-calibration is done on C2, array's focus shifts to C2, closer to T than C1.  The more-sophisticated calibration estimates the solution of self-calibration on a hypothetical calibrator~C2$^\prime$ from the solution on C2.  Array's focus finally shifts to C2$^\prime$, closer to T than C2.  

The key of the BPR is the presence of suitable C2 at a position less half the distance to T than that to C1.  Without such a suitable C2, a BPR image will be worse than that of normal phase referencing.  However, even a distant C2 may help the more-sophisticated calibration, if the sources align straightly.  

The key of the more-sophisticated calibration is the linearity of the phases $\Delta\phi_\mathrm{atmo}$, $\Delta\phi_\mathrm{geo}$, and $\Delta\phi_\mathrm{pos}$.  The linearity around sources have been already assumed in the method of ATMCA \citep{Fomalont&Kogan2005}, which uses strong calibrators around a target to determine the phase gradient.  The best performance can be obtained by both the more-sophisticated calibration and the ATMCA when sources align straightly.  Even in case of misaligned source allocation, the calibration of ATMCA can be done by estimation of a two-dimensional phase gradient using three or more calibrators.  Because of the same assumption, the more-sophisticated calibration also can do it.  Array's focus can be established exactly at T, when $\overrightarrow{\rm C_1 C_2^{\ \prime}} + \overrightarrow{\rm C_1 C_3^{\ \prime}}$
, i.e. $r_{12} \overrightarrow{\rm C_1 C_2} + r_{13} \overrightarrow{\rm C_1 C_3}$, $= \overrightarrow{\rm C_1 T}$.  Such a calibration table is derived from the two solution tables produced by self-calibrations on C2 and C3, in addition to that on C1.  Not only bright calibrators but also weak ones are usable as secondary calibrators, when the BPR is applied before the more-sophisticated calibration or the ATMCA.  Alternatively, with only one secondary calibrator, a good calibration can be achieved if assuming that the phase gradient is in elevation direction because the residual phase error is mostly tropospheric origin.  This is supported by an option of the ATMCA.  In such a process, observers should separate $\Delta\phi_\mathrm{atmo}$ from the other terms that would interact with the determination of the phase gradient.  \citet{Fomalont&Kogan2005} recommend observing for more than 3~hours to find the true positions of calibrators, in order to make $\Delta\phi_\mathrm{pos} \approx 0$.  The term $\Delta\phi_\mathrm{geo}$, however, would be harmful in arrays with poorly position-determined spacecraft and/or antennas.

The BPR can make very weak sources available as calibrators.  In normal phase-referencing a signal-to-noise ratio of more than $\sim$5 in half of a fast-switching period is required to obtain phase, delay, delay-rate solutions, while in the BPR a signal-to-noise ratio of $\sim$5 to obtain only a phase solution in about a quarter of a pair-swapping period is acceptable.  This means that we can actually use calibrators more than a~few times weaker than conventional calibrators.  Consequently, the BPR brings lots of benefits, for example: (1)~observers will be given more opportunities to make phase-referencing successfully at short centimeter and millimeter bands ($\sim$22--86~GHz), where targets normally have little chance of being accompanied by bright calibrators with a small separation angle.  (2)~Phase-referencing is available even to poorly sensitive or poorly position-determined antennas such as a space-VLBI.  (3)~Astrometry can be made between two weak sources, one of which has been self-calibrated.  (4)~Observers can easily obtain idealized calibrator arrangements on the sky plane for the subsequent more-sophisticated calibration or the ATMCA, because of much larger surface number density in the faint-source sky.  

Practically, prior phase-referencing observations will be needed to find suitable calibrators around the target.  We suggest that candidates are selected from flat-spectrum sources in the catalogs of the FIRST survey and so on, like an approach of `a VLBA survey of flat-spectrum FIRST sources~\citep{Ulvestad_etal.1999}.'

The BPR and the subsequent more-sophisticated calibration are universal-designed.  There is no requirement of upgrades for hardware or software.  All observers have to do is scheduling of fast-switching with less frequent pair-swapping, and data reduction with bigradient calibration processes, which can be done only with the AIPS and editting solution tables.  Even with antennas whose positions have not been well-determined, or even with a correlator without precise atmospheric models, very weak targets can be detected because of a much smaller separation from C2 or C2$^\prime$ than that from C1.  A-priori calibration by geodetic-like observations (e.g., \cite{Mioduszewski&Kogan2004}) should bring independent improvements.  Therefore, the combination of such prior calibrations and our method should achieve extremely high-quality phase-referencing.

\section{Summaries}\label{section:summary}
The bigradient phase referencing~(BPR) allows us to utilize weak calibrators, one of which may be located at a position very closer to a target.  The subsequent more-sophisticated calibration makes array's focus to shift to a hypothetical point much closer to a target.  Thanks to much smaller separation angle, the phase-referencing quality of a target image will be dramatically improved.  We describe the theory of the method and expected errors.  For relative astrometry, observer must use identical phase-tracking centers every epoch and the first self-calibrations on the calibrators using the structure models at the phase-tracking centers.  Our demonstrative observation tests with strong sources~(3C~345, NRAO~512, and DA~426) have shown its capabilities.  Image dynamic ranges have been dramatically improved by a factor of about six, compared to normal phase-referencing.  An astrometric reproducibility is 62$\pm$50~$\mu$as between our two epochs in the most calibrated cases.  The observation case of a weak calibrator, which is originally supposed in the BPR, will be reported in a future paper.  

\bigskip


The auther wishes to thank the JVN team for observing assistance and support.  We are grateful to K.~Fujisawa for comments on an early draft of the manuscript, and to N.~Kawaguchi, M.~Honma, and H.~Kobayashi for valuable discussions.  We thank the anonymous referee for excellent suggestions and thoughtful comments.  The JVN project is led by the National Astronomical Observatory of Japan~(NAOJ) that is a branch of the National Institutes of Natural Sciences~(NINS), Hokkaido University, Gifu University, Yamaguchi University, and Kagoshima University, in cooperation with Geographical Survey Institute~(GSI), the Japan Aerospace Exploration Agency~(JAXA), and the National Institute of Information and Communications Technology~(NICT).


\clearpage

\clearpage

\end{document}